\documentclass[12pt]{article}
\usepackage[dvips]{graphicx}
\def\be{\begin{equation}}					 
\def\ee{\end{equation}}
\def\ber{\begin{eqnarray}}
\def\eer{\end{eqnarray}}	
\def\dint{\mathop{\intop\kern-0.5em\intop}}
\def\ovc#1{\displaystyle\mathop{#1}^{\kern0.2em\circ}}

\begin{document}
\vspace*{1cm}
\begin{center}
{\Large \bf Three-Body Forces in Multiparticle Dynamics:\\[1ex] New
Sites of Fundamental Dynamics.}\footnote{The talk presented at XXXII
International Symposium on Multiparticle Dynamics, Alushta, Crimea,
Ukraine, September 7--12,
2002.} 

\vspace{4mm}

{\large A.A. Arkhipov\\
{\it State Research Center ``Institute for High Energy Physics" \\
 142280 Protvino, Moscow Region, Russia}}\\
\end{center}

\vspace{4mm}
\begin{abstract}
A manifestation of the three-body forces in multiparticle
dynamics is discussed. The minireview of our recent results has been
presented.
\end{abstract}

\section{Introduction}
Three-body forces are fundamental forces which take place in the
multiparticle systems where the number of particles is greater than
two. The simplest one of such systems is a three-particle system.
However, it's clear that in any multiparticle system there are
two-body or pair interactions between particles and the question
arises how could we separate three-body forces from two-body ones. Of
course, a definition of three-body forces depends on a formalism
which one uses. It turns out that single-time formalism in Quantum
Field Theory is a useful tool to make it.

Using the LSZ or Bogoljubov reduction formulae in Quantum Field
Theory we can easily obtain the following cluster structure
for $3\rightarrow 3$ scattering amplitude 
\be
{\cal F}_{123} = {\cal F}_{12} + {\cal F}_{23} + {\cal F}_{13} + 
{\cal F}_{123}^C, \label{cluster}
\ee
where ${\cal F}_{ij} , (i,j = 1,2,3,i\not= j)$ are $2
\rightarrow
2$
scattering  amplitudes, ${\cal F}_{123}^C$ is called the connected
part of the $3 \rightarrow 3$ scattering amplitude.

\vskip0.4cm
\begin{center}
\begin{picture}(370,30)
\thicklines
\put(0,0){\line(1,0){60}}
\put(0,30){\line(1,0){60}}
\put(30,15){\circle{30}}
\put(0,15){\line(1,0){15}}
\put(45,15){\line(1,0){15}}

\put(63,12){\mbox{=}}

\put(78,0){\line(1,0){60}}
\put(78,15){\line(1,0){60}}
\put(108,22,5){\circle{15}}
\put(78,30){\line(1,0){60}}

\put(140,12){\mbox{+}}

\put(152,0){\line(1,0){60}}
\put(152,15){\line(1,0){60}}
\put(152,30){\line(1,0){60}}
\put(182,7.5){\circle{15}}

\put(215,12){\mbox{+}}

\put(230,0){\line(1,0){60}}
\put(230,30){\line(1,0){60}}
\put(260,15){\circle{30}}
\put(260,15){\oval(40,20)[b]}
\put(230,15){\line(1,0){10}}
\put(280,15){\line(1,0){10}}

\put(300,12){\mbox{+}}

\put(315,0){\line(1,0){60}}
\put(315,30){\line(1,0){60}}
\put(345,15){\circle{30}}
\put(315,15){\line(1,0){15}}
\put(360,15){\line(1,0){15}}

\put(335,7){\mbox{\huge C}}

\end{picture}
\end{center}
In the framework of the single-time formalism in Quantum Field Theory
developed in \cite{1} we construct the $3 \rightarrow 3$ off energy
shell scattering amplitude $T_{123}(E)$ with the same (cluster)
structure 
\be
T_{123}(E) = T_{12}(E) + T_{23}(E) + T_{13}(E) +
T_{123}^C(E).\label{st_cluster}
\ee
The three particle interaction quasipotential $V_{123}(E)$ is related
to the off energy shell $3 \rightarrow 3$ scattering amplitude
$T_{123}(E)$ by the Lippmann-Schwinger transformation
(LS-transformation)
\be
T_{123}(E) = V_{123}(E) + V_{123}(E)G_0(E)T_{123}(E).
\label{quasipotential}
\ee
There exists the same LS-transformation between two particle
interaction quasi\-potentials $V_{ij}$ and off energy shell $2
\rightarrow 2$ scattering amplitudes $T_{ij}\ \ (i,j = 1,2,3,i\not=
j)$
\[
T_{ij}(E) = V_{ij}(E) + V_{ij}(E)G_0(E)T_{ij}(E).
\]
In Quantum Field Theory a three particle interaction quasipotential
has the structure \cite{2}
\be
V_{123}(E) = V_{12}(E) + V_{23}(E) + V_{13}(E) +
V_0(E),\label{V_cluster}
\ee
where $V_0(E)$ is the three-body forces quasipotential; it represents
the defect of three particle interaction quasipotential over the sum
of two particle quasipotentials and describes the true
three-body interactions. Three-body forces quasipotential is an
inherent connected part of total three particle interaction
quasipotential which cannot be represented by the sum of pair
quasipotentials. The three-body forces could be seen even in the
perturbation theory expansions. However, and we would like to stress
it, formulae (\ref{cluster}--\ref{V_cluster}) provide a constructive
definition of the three-body forces in the framework of local Quantum
Field Theory without having the use of perturbation theory.

The three-body forces scattering amplitude is related to the
three-body forces quasipotential by the LS-transformation
\be
T_0(E) = V_0(E) + V_0(E)G_0(E)T_0(E).\label{3-forces}
\ee
It should be pointed out once more that the three-body forces appear
as a result of consistent consideration of the three-body problem in
the framework of local Quantum Field Theory.

\section{Global analyticity of the three-body forces}

Let us introduce the following useful notations:
\be
<p'_1 p'_2 p'_3\vert S - 1\vert p_1 p_2 p_3> =
2\pi i\delta ^4(\sum_{i=1}^{3}p'_i-\sum_{j=1}^{3}p_j)
{\cal F}_{123}(s;{\hat e}',\hat e), \label{8}
\ee
\[
s = (\sum_{i=1}^{3}p'_i)^2 = (\sum_{j=1}^{3}p_j)^2.
\]
${\hat e}', \hat e \in S_5$ are two unit vectors on five-dimensional
sphere describing the configuration of three-body system in initial
and final states (before and after scattering). We will denote the
quantity $T_0$ restricted on the energy shell as
\[
T_0\mid_{on\, energy\, shell}\, = {\cal F}_0.
\]
The unitarity condition for the quantity ${\cal F}_0$ with 
account for the introduced notations can be written in form
\cite{3,4}
\be
Im{\cal F}_0(s;{\hat e}',\hat e) = \pi A_3(s)\int d\Omega _{5}({\hat
e}''){\cal F}_0(s;{\hat e}',{\hat e}'')
\stackrel{*}{\cal F}_0(s;\hat e,{\hat e}'') + H_0(s;{\hat e}',\hat
e),
\label{9}
\ee

\[
Im{\cal F}_0(s;{\hat e}',\hat e)\equiv\frac{1}{2i}\left[{\cal
F}_0(s;{\hat
e}',\hat e)
-\stackrel{*}{\cal F}_0(s;\hat e,{\hat e}')\right],
\]
where
\[
A_3(s) = {\Gamma}_3(s)/S_5 ,
\]
${\Gamma}_3(s)$ is the three-body phase-space volume, $S_5$ is the
volume of unit five-dimensional sphere, $H_0$ defines the
contribution of all inelastic channels emerging due to three-body
forces.

Let us introduce a special notation for the scalar product of two
unit vectors 
${\hat e}'$ and $\hat e$
\be
\cos \omega = {\hat e}^{'}\cdot \hat e. \label{10}
\ee
The other notation for the three-body forces scattering
amplitude will be used as well 
\[
{\cal F}_0(s;{\hat e}',\hat e) = {\cal F}_0(s;\eta,\cos\omega),
\]
where all other variables are denoted through $\eta$. Now we are able
to formulate our basic assumption concerning the
analytical properties of the three-body forces scattering
amplitude 
\cite{3,4}.

We will assume that for physical values of the variable $s$ and fixed 
values of  $\eta$ the amplitude ${\cal F}_0(s;\eta,\cos\omega)$ is an 
analytical function of the variable $\cos\omega$ in the ellipse
$E_0(s)$ 
with the semi-major axis
\be
z_0(s) = 1 + \frac{M_0^2}{2\,\Pi^2(s)} \label{11}
\ee
and for any $\cos\omega \in E_0(s)$ and physical values of $\eta$ it
is polynomially bounded in the variable $s$. $M_0$ is some constant
having mass dimensionality.

Such analyticity of the three-body forces amplitude was called a
global one. The global analyticity may be considered as a direct
geometric generalization of the known analytical properties of
two-body scattering amplitude strictly proved in the local quantum
field theory. At the same time the global analyticity
results in the generalized asymptotic bounds. 
\begin{center}
$\framebox{\bf GLOBAL ANALYTICITY}\enspace \& \enspace \framebox{\bf
UNITARITY}$\\[1ex]
$\Downarrow$\\[1ex]
$\framebox{\bf GENERALIZED ASYMPTOTIC BOUNDS}$
\end{center}
For example, the generalized asymptotic bound for
$O(6)$-invariant three-body forces scattering amplitude looks like
\cite{3,4}
\begin{equation}
Im\,{\cal F}_0(s;\cos\omega=1) \leq \mbox{Const}\, s^{3/2} 
\bigl(\frac{\ln s/s'_0}{M_0}\bigr)^5 = \mbox{Const}\,
s^{3/2}R_0^5(s),
\label{12}
\end{equation}
where $R_0(s)$ is the effective radius of the three-body forces
\be
R_0(s) = \frac{\Lambda_0}{\Pi(s)} = \frac{r_0}{M_0}\ln
\frac{s}{s'_0},
\quad \Pi(s) = \frac{\sqrt{s}}{2},\quad s\rightarrow \infty,
\label{reff}
\ee
$r_0$ is defined by the power of growth of the amplitude ${\cal F}_0$  
at high energies \cite{4}, $M_0$ defines the semi-major axis of the
global analyticity ellipse (\ref{11}), ${\Lambda}_0$ is the effective
global orbital momentum, $\Pi(s)$ is the global momentum of
three-body system, $s'_0$ is a scale defining a range of unitarity
saturation of three-body forces.
 
If we want to have a possibility for an experimental verification of
generalized asymptotic bound (\ref{12}) we have to establish a
connection between the three-body forces scattering  amplitude and
the experimentally measurable quantities. In fact, we found the
connection of the three-body forces scattering amplitude with the
experimentally measurable quantity which is the total cross section
in scattering from deuteron \cite{5,6}, and  the relation of the
three-body forces scattering amplitude to one-particle inclusive
cross sections was established as well \cite{7}. 

Now we shall briefly sketch the basic results of our analysis. 

\section{Three-body forces in single diffraction dissociation}

The formula relating one-particle inclusive cross-section with the
imaginary part of the three-body forces scattering amplitude looks
like \cite{6,7}
\be 
\fbox{$\displaystyle 2E_N(\vec{\Delta})\frac{{d\sigma}_{hN\rightarrow
NX}}{d\vec{\Delta}}(s,\vec{\Delta}) = 
- \frac{(2\pi)^3}{I(s)}
Im{\cal F}_0^{scr}(\bar s;-\vec{\Delta}, \vec{\Delta}, \vec q; 
\vec{\Delta}, -\vec{\Delta}, \vec q\,)$}\,, \label{13}
\ee 
\vspace{3mm}
\[
Im{\cal F}_0^{scr}(\bar s;-\vec{\Delta}, \vec{\Delta}, \vec q; 
\vec{\Delta}, -\vec{\Delta}, \vec q\,) = Im{\cal F}_0(\bar
s;-\vec{\Delta}, \vec{\Delta}, \vec q; 
\vec{\Delta}, -\vec{\Delta}, \vec q\,)-
\]
\[
-
4\pi\int d\vec{\Delta}'\frac{\delta\left[E_N(\vec{\Delta} -
\vec{\Delta}') + \omega_h(\vec q+\vec{\Delta}') - E_N(\vec{\Delta}) -
\omega_h(\vec q)\right]}{2\omega_{h}(\vec q + \vec{\Delta}')2E_N
(\vec{\Delta} - \vec{\Delta}')}\times
\]
\be
Im{\cal F}_{hN}(\hat s; \vec{\Delta}, \vec q; 
\vec{\Delta}-\vec{\Delta}', \vec q + \vec{\Delta}'\,)Im{\cal
F}_0(\bar s;-\vec{\Delta}, \vec{\Delta}-\vec{\Delta}', \vec q +
\vec{\Delta}'; 
\vec{\Delta}, -\vec{\Delta}, \vec q\,), \label{14}
\ee 
\vspace{3mm}
\[
E_N(\vec{\Delta})=\sqrt{{\vec{\Delta}}^2+M_N^2},\quad
\omega_h(\vec q)=\sqrt{{\vec q}\,^2+m_h^2},\quad
I(s) = 2{\lambda}^{1/2}(s,m_h^2,M_N^2),
\]
\be
\hat s = \frac{\bar s +
m_h^2 - 2M_N^2}{2},\quad
\bar s = 2(s + M_N^2) - M_X^2,\quad t = - 4{\vec\Delta}^2. \label{15}
\ee
I'd like to draw attention to the minus sign in the R.H.S. of
Eq.~(\ref{13}). Because we have a positive physical quantity in the
L.H.S. this means that second term in the R.H.S. of
Eq.~(\ref{14}) has to be dominated over the first one. It's really
true. The simple model for the three-body forces
\[
Im\,{\cal F}_0(s; \vec p_1, \vec p_2, \vec p_3; \vec q_1, \vec q_2,
\vec q_3)
= f_0(s)
\exp \Biggl\{-\frac{R^2_0(s)}{4} \sum^{3}_{i=1} (\vec p_i-\vec
q_i)^2\Biggr\},
\]
where $f_0(s)$, $R_0(s)$ are model parametric functions of $s$,
gives the following result for the one-particle inclusive cross
section  in the region of diffraction dissociation
\[
\frac{s}{\pi}\frac{d\sigma_{hN\rightarrow NX}}{dtdM_X^2} =
\frac{(2\pi)^3}{I(s)}\chi(\bar s)Im{\cal F}_0(\bar
s;-\vec{\Delta}, \vec{\Delta}, \vec q; 
\vec{\Delta}, -\vec{\Delta}, \vec q\,)
\]
\be
= \frac{(2\pi)^3}{I(s)}\chi(\bar s)f_0(\bar
s)\exp\Biggl[\frac{R_0^2(\bar s)}{2}t\Biggr],\label{16}
\ee
where
\[
\chi(\bar s) = \frac{\sigma^{tot}_{hN}({\bar s}/2)}{2\pi[B_{hN}({\bar
s}/2) + R_0^2(\bar s)]} -1.
\]
The function $\chi$ has a clear physical meaning: This function
originates from initial and final states interactions and  describes
the shadowing (eclipsing) effect or the effect of screening the
three-body forces by two-body ones \cite{7}.

If we take the usual parametrization for one-particle inclusive
cross-section in the region of diffraction dissociation
\be
\frac{s}{\pi}\frac{d\sigma}{dtdM_X^2} = A(s.M_X^2)\exp[b(s,M_X^2)t],
\label{18}
\ee 
then one obtains for the quantities $A$ and $b$
\be
A(s,M_X^2) = \frac{(2\pi)^3}{I(s)}\chi(\bar s)f_0(\bar
s),\quad 
b(s,M_X^2) = \frac{R_0^2(\bar s)}{2} \label{19}.
\ee
Eq.~(\ref{19}) displays a remarkable fact: 
\vskip 0.1 true in
\framebox[153mm]{\parbox{150mm}{{\bf the effective radius of
three-body forces is  related to the slope of diffraction cone for
inclusive diffraction dissociation processes in the same way as the
effective radius of two-body forces is related to the slope of
diffraction cone in elastic scattering processes.}}}\\[1ex]
\noindent
Moreover, from the expressions
\[
R_0(\bar s) = \frac{r_0}{M_0} \ln \bar s/s'_0,\quad \bar s =
2(s+M_N^2) - M_X^2
\]
it follows that 
\vskip 0.1 true in
\framebox[153mm]{\parbox{150mm}{\bf the slope of diffraction cone for
inclusive diffraction dissociation processes at fixed energy
decreases  with the growth  of missing mass and increases with the
growth  of energy at fixed value of missing mass.}}\\[1ex]
This property agrees well qualitatively 
with the experimentally observable picture. Actually, we have even a 
more remarkable fact: Shrinkage or narrowing of diffraction cone for
inclusive diffraction dissociation processes with the growth of
energy at a fixed missing mass and widening of this cone with the
growth of missing mass at a fixed energy is of universal character.
As it follows from Eq.~(\ref{16}) this property is the consequence of
the fact that the one-particle inclusive cross-section up to flux
factor depends on the variables $s$ and $M_X^2$ via one variable
$\bar s$ which is a linear combination of $s$ and $M_X^2$. This
peculiar ``scaling" is the  manifestation of $O(6)$-symmetry of the
three-body forces. It would be very desirable to
experimentally study this new scaling law related to the symmetry of
the new fundamental (three-body) forces.

There is a very important relation
\be
\fbox{$\displaystyle A(s,M_X^2) = \frac{\bar s M_N
[R_0^2(\bar
s)+R_d^2]^{3/2}}{(2\pi)^{3/2}I(s)}\delta\sigma^{inel}(\bar s)
$}\,.\label{21}
\ee
Eq.~(\ref{21}) establishes a deep connection of inelastic shadow
correction in scattering from deuteron with one-particle inclusive
cross-section. This relation for one's turn allowed us to express the
inelastic shadow correction via a total single diffractive
dissociation cross-section \cite{8}. This will be shown in the next
section.

\section
{Three-body forces in scattering from deuteron}

We have applied our approach to study a shadow dynamics in scattering
from deuteron in some details. In this way a
new simple formula for the shadow corrections to the total
cross-section in scattering from deuteron has been derived and new
scaling characteristics with a clear physical interpretation have
been established. Here we shall briefly sketch the basic results of
our analysis of high-energy particle scattering from deuteron. As has
been shown in \cite{8}, the total cross-section in the scattering
from deuteron can be expressed by the formula
\[
\sigma_{hd}^{tot}(s) = \sigma_{hp}^{tot}(s)
+\sigma_{hn}^{tot}(s) -
\delta\sigma(s), 
\]
where $\sigma_{hd}, \sigma_{hp}, \sigma_{hn}$  are the total
cross-sections in scattering from deuteron, proton and neutron, 
\be
\delta\sigma(s) = \delta\sigma^{el}(s)
+\delta\sigma^{inel}(s)=2\sigma^{el}(s)a^{el}(x_{el}) +
2\sigma_{sd}^{ex}(s)a^{inel}(x_{inel}) ,\label{46}
\ee
\[
\sigma^{el}(s) \equiv \frac{\sigma_{hN}^{tot\,2}(s)}{16\pi
B_{el}(s)},\quad a^{el}(x_{el}) = \frac{x^2_{el}}{1+x^2_{el}},
\quad
x^2_{el} \equiv \frac{2B_{el}(s)}{R_d^2} =
\frac{R_2^2(s)}{R_d^2},
\]
\[
a^{inel}(x_{inel}) = \frac{x^2_{inel}}{(1+x^2_{inel})^{3/2}}, \quad
x^2_{inel} \equiv \frac{R_3^2(s)}{R_d^2} =
\frac{2B_{sd}(s)}{R_d^2},\quad B_{sd}(s)\equiv
b(s,M_X^2)|_{M_X^2=2M_N^2}.
\]
The total single diffractive dissociation cross-section
$\sigma_{sd}^{ex}(s)$ is defined by the following equation \cite{8}
\be
\sigma_{sd}^{\varepsilon}(s) =
\pi\int_{M_{min}^2}^{\varepsilon
s}\frac{dM_X^2}{s}\int_{t_{-}(M_X^2)}^{t_{+}
(M_X^2)} dt \frac{d\sigma}{dtdM_X^2}, \label{47}
\ee
where
\be
\varepsilon=\varepsilon^{ex} = \sqrt{2\pi}/2M_N R_d,\label{ex}
\ee
and we supposed that
$\sigma_{hp}^{tot}=\sigma_{hn}^{tot}=\sigma_{hN}^{tot}$ and
$B_{el}^{hp}=B_{el}^{hn}=B_{el}$ at high energies. The first term in
the R.H.S. of Eq.~(\ref{46}) generalizes the known
Glauber correction
\[
\delta\sigma^{el}(s)=\delta\sigma_G(s)=\frac{\sigma_{hN}^{tot\,2}(s)}
{4\pi R_d^2},\quad  x^2_{el}<<1,
\]
but the second term in the R.H.S. of Eq.~(\ref{46}) is totally new
and comes from the contribution of the three-body  forces to the
hadron-deuteron total cross section. 

The expressions for the shadow corrections have quite a 
transparent physical meaning, both elastic $a^{el}$ and
inelastic $a^{inel}$ scaling functions related to elastic and
inelastic parts of the total shadow correction have a clear physical
interpretation \cite{9}. The function $a^{el}$  measures out a
portion of elastic rescattering events among of all the events during
the  interaction of an incident particle with a deuteron as a whole,
and this function attached to the total probability of elastic
interaction of an incident particle with a separate nucleon in a
deuteron. Correspondingly, the function $a^{inel}$ measures out a
portion of inelastic events of inclusive type among of all the events
during  the interaction of an incident particle with a deuteron as a
whole, and this function attached to the total probability of
single diffraction dissociation  of an incident particle on a
separate nucleon in a deuteron. The scaling variables $x_{el}$ and
$x_{inel}$ have quite a clear physical meaning too. The dimensionless
quantity $x_{el}$ characterizes the effective distances measured in
the units of ``fundamental length", which the deuteron size is, in
elastic interactions, but the similar quantity $x_{inel}$
characterizes the effective distances measured in the units of the
same ``fundamental length" during inelastic interactions.

The functions $a^{el}$ and $a^{inel}$ have a different
behaviour: $a^{el}$ is a monotonic function while $a^{inel}$ has the 
maximum at the point $x^{max}_{inel}=\sqrt{2}$ where 
$a^{inel}(x^{max}_{inel})=2/3\sqrt{3}$. The existence of the maximum
in the function $a^{inel}$ results an interesting physical
effect of weakening the inelastic eclipsing (screening) at
superhigh energies. The energy $s_m$ at the maximum of $a^{inel}$ can
be calculated from the equation $R_3^2(s_m)=2 R_d^2$, and one obtains
in this way: $\sqrt{s_m}=9.01\,10^8\,GeV=901\,PeV$ \cite{8,9}. 

From geometrical point of view, if we would consider the standard
Glauber correction as the shadowing where one two-dimensional sphere
is in eclipse with the other two-dimensional sphere, then the
inelastic screening and the effect of its weakening should be
considered as the shadowing where a five-dimensional sphere ``casts"
its shadow on a two-dimensional sphere. So, we really see here the
true play of the spheres. 

It is very important that formula (\ref{46}) contains only
experimentally measured quantities.
Therefore we can verify the new structure for the shadow corrections
in elastic scattering from deuteron using the existing experimental
data on proton-deuteron and antiproton-deuteron total cross sections
and experimental data on single diffractive dissociation in
scattering from nucleon. The results of comparison are shown in
Figs.~1-2 extracted from paper \cite{8}. 

We would like to emphasize that in the fit to the data on
antiproton-deuteron total cross sections our global fit to the data
on antiproton-proton total cross section \cite{10} has been used (see
next section), and
$R_d^2$ was considered as a single free fit parameter. After that a
comparison with the data on proton-deuteron total cross sections has
been made without any free parameters: $R_d^2$ was fixed by the
previous fit to the data on antiproton-deuteron total cross sections,
our fit yielded $R_d^2=66.61\pm 1.16\,GeV^{-2}$, and our global fit
to the data on proton-proton total cross section \cite{10} has been
used as well (see next section). If we take into account the
latest experimental value for the deuteron matter radius
$r_{d,m}=1.963(4)\,fm$ \cite{11} then we can find that the fitted
value for the $R_d^2$ satisfies with a good accuracy the equality
\cite{8,9}
\be
R_d^2 = \frac{2}{3}r^2_{d,m},\quad  (r^2_{d,m} = 3.853\,\mbox{fm}^2 =
98.96\,\mbox{GeV}^{-2}).\label{rd}
\ee

\section
{Global structure of $\bar pp$ and $pp$ total cross
sections}

Recently \cite{7,10} a simple theoretical formula describing the
global structure of $pp$ and  $p\bar p$ total cross-sections in the
whole range of energies available up today has been derived by an
application of single-time formalism in QFT and general theorems a
l\`a Froissart. The fit to the experimental data with the formula was
made, and it was shown that there is a very good correspondence of
the theoretical formula to the existing experimental data obtained at
the accelerators. Moreover, it turned out \cite{12} there is a very
good correspondence of the theory to all existing cosmic ray
experimental data  as well: The predicted values for
$\sigma^{tot}_{pp}$ obtained from theoretical description of all
existing accelerators data are completely compatible with the values
obtained from cosmic ray experiments \cite{13}. The global structure
of (anti)proton-proton total cross section is shown in Figs.~3-4
extracted from papers \cite{10,12}.

The theoretical formula describing the global structure of
(anti)proton-proton total cross section has the following
structure \cite{10}

\be
\sigma_{(\bar p)pp}^{tot}(s) = \sigma^{tot}_{asmpt}(s) 
\left[1 + \chi_{(\bar p)pp}(s)\right],\label{42}
\ee
where
\be
\sigma^{tot}_{asmpt}(s) = 2\pi\left[B_{el}(s) +
(1-\beta)R_3^2(s)\right] = \left[42.0479 + 1.7548
\ln^2(\sqrt{s}/20.74)\right](mb),\label{43}
\ee
\[
B_{el}(s) = R_2^2(s)/2 = \left[11.92 + 0.3036
\ln^2(\sqrt{s}/20.74\right](GeV^{-2}),
\]
\[
R^2_3(s)|_{\beta<<1} = \left[0.40874044 \sigma^{tot}_{asmpt}(s)(mb) -
B_{el}(s)\right](GeV^{-2}) =
\]
\be
=\left[5.267+0.4137\ln^2\sqrt{s}/20.74\right](GeV^{-2}),\label{44}
\ee
\[
\beta = \frac{x^2_{inel}}{4(1+x^2_{inel})},\quad
x^2_{inel}=\frac{R_3^2(s)}{R_d^2}=\frac{2B_{sd}(s)}{R_d^2},
\]
$B_{el}(s)$ is the slope of nucleon-nucleon differential elastic
scattering cross section, $R_2(s)$ is the effective radius of
two-nucleon forces, $R_3(s)$ is the effective radius of three-nucleon
forces, $R_d$  characterizes the internucleon distance in a deuteron,
the functions $\chi_{(\bar p)pp}(s)$ describe low-energy parts of
(anti)proton-proton total cross sections and asymptotically tend to
zero at $s\rightarrow\infty$ (see details in the 
original paper \cite{10}) 
\be
\chi_{\bar pp}(s) = \frac{c}{\sqrt{s-4m^2_N}R^3_0(s)}
\left(1+\frac{d_1}{\sqrt{s}} +
\frac{d_2}{s} + \frac{d_3}{s^{3/2}}\right),\label{chipbarp}
\ee
\[
d_1 = (-12.12\pm 1.023)GeV,\quad d_2 = (89.98\pm 15.67)GeV^2,
\]
\[
d_3 = (-110.51\pm 21.60)GeV^3,\quad c = (6.655\pm 1.834)GeV^{-2}.
\]
and
\be
\chi_{pp}(s) = \left(\frac{c_1}{\sqrt{s-4m^2_N}R^3_0(s)} -
\frac{c_2}{\sqrt{s-s_{thr}}R^3_0(s)}\right)\left(1 +
d(s)\right)\vert_{s>s_{thr}} + Res(s),\label{chipp}
\ee
\[
c_1 = (192.85\pm 1.68)GeV^{-2},\quad c_2 = (186.02\pm 1.67)GeV^{-2},
\]
\[
d(s) = \sum_{k=1}^{8}\frac{d_k}{s^{k/2}},\qquad Res(s) =
\sum_{i=1}^{N}\frac{C_R^i s_R^i
{\Gamma_R^i}^2}{\sqrt{s(s-4m_N^2)}[(s-s_R^i)^2+s_R^i{\Gamma_R^i}^2]},
\]
\be
\fbox{$\displaystyle s_{thr}=(3.5283\pm
0.0052)GeV^2$}\,.\label{thresh}
\ee\\
For the numerical values of the parameters $d_i (i=1,...8)$ see
original paper \cite{10}. 

The formula (\ref{42}) represents the total cross section in a
factorized form. One factor describes high energy
asymptotics of total cross section and it has the universal energy
dependence predicted by the general theorems in local Quantum Field
Theory (Froissart theorem). The other factor is responsible for the
behaviour of total cross section at low energies and it has a
complicated resonance structure.  However this factor
has also the universal asymptotics at elastic threshold. It is a
remarkable fact that the low energy part of total cross section has
been derived by application of the generalized Froissart theorem for
a three-body forces scattering amplitude. 

The nontrivial feature of the formula for the proton-proton total
cross section is the presence of the new ``threshold"
$s_{thr}=3.5283\, GeV^2$ which is near the elastic one. Moreover,
Eq.~(\ref{43}) shows that geometrical scaling in a naive form
$\sigma^{tot}_{asmpt}(s) = \mbox{Const}B_{el}(s)$ is not valid.
However, from  Eq.~(\ref{43}) it follows the generalized  geometrical
scaling which looks like 
\be
\sigma^{tot}_{asmpt}(s) = 2\pi B_{el}(s)[1 +
2\gamma (1-\beta )],\label{gm-scale}
\ee
where $\beta$ is defined above and
\[
\gamma = \frac{R_3^2(s)}{2B_{el}(s)}=\frac{R_3^2(s)}{R_2^2(s)} =
\frac{B_{sd}(s)}{B_{el}(s)}.
\]

Some information concerning the diproton resonances is collected in
Table 1. The positions of resonances and their widths, listed in
Table 1, were fixed in our fit, and only relative contributions of
the resonances $C_R^i$ have been considered as free fit parameters.
Fitted parameters $C_R^i$ obtained by the fit are listed  in Table 1
too. 

Our fitting curve concerning low-energy region is shown in Fig.~5. We
also plotted in Fig.~6 the
resonance structure of proton-proton total cross section at low
energies without the experimental points but with dashed line
corresponding the ``background" where all resonances are switched
off. As it is seen from this Figure there is a clear signature for
the diproton resonances. We may conclude that the diproton resonances
are confirmed by the data set for proton-proton total cross section
at low energies from statistical point of view by the good fit
\cite{16}.

From the global structure of proton-proton total cross-section it
follows that the new ``threshold", which is near the elastic one,
looks like a manifestation of a new unknown particle:

\be
\sqrt{s_{thr}} = 2 m_p + m_{\cal L},\qquad m_{\cal L} =
1.833\,MeV.\label{L}
\ee

\begin{center}
Table 1. Diproton resonances.

\vspace{5mm}
\begin{tabular}{|l|c|c|r|}\hline   
$ m_R(MeV) $ & $\Gamma_R(MeV) $ & \mbox{Refs.} & $C_R(GeV^2)$  \\
\hline     
$ 1937\pm 2 $ & $ 7\pm 2 $ & $\cite{14}$ & $ 0.058\pm 0.018 $ \\ 
$ 1947(5)\pm 2.5 $ & $ 8\pm 3.9 $ & $\cite{15}$ & $ 0.093\pm 0.028
$\\ 
$ 1955\pm 2 $ & $ 9\pm 4 $ & $\cite{14}$ & $ 0.158 \pm 0.024 $ \\
$ 1965\pm 2 $ & $ 6\pm 2 $ & $\cite{14}$ & $ 0.138 \pm 0.009 $ \\ 
$ 1980\pm 2 $ & $ 9\pm 2 $ & $\cite{14}$ & $ 0.310 \pm 0.051 $ \\
$ 1999\pm 2 $ & $ 9\pm 4 $ & $\cite{14}$ & $ 0.188\pm 0.070 $ \\ 
$ 2008\pm 3 $ & $ 4\pm 2 $ & $\cite{14}$ & $ 0.176 \pm 0.050 $ \\ 
$ 2027\pm ? $ & $ 10 - 12 $ &$\cite{16}$ & $ 0.121\pm 0.018 $ \\ 
$ 2087\pm 3 $ & $ 12\pm 7 $ & $\cite{14}$ & $ -0.069\pm 0.010 $ \\ 
$ 2106\pm 2 $ & $11\pm 5 $ & $\cite{14}$ & $-0.232 \pm 0.025 $ \\ 
$ 2127(9)\pm 5 $ & $ 4\pm 2 $ & $\cite{14}$ & $ -0.222\pm 0.056 $ \\ 
$ 2180(72)\pm 5 $ & $ 7\pm 3 $ & $\cite{14}$ & $ 0.131\pm 0.015 $ \\ 
$ 2217\pm ? $ & $ 8 - 10 $ &$\cite{16}$ & $ 0.112\pm 0.031 $ \\ 
$ 2238\pm 3 $ & $22\pm 8 $ & $\cite{14}$ & $ 0.221 \pm 0.078 $ \\ 
$ 2282\pm 4 $ & $24\pm 9 $ & $\cite{14}$ & $ 0.098 \pm 0.024 $ \\
\hline
\end{tabular}
\end{center}
\vspace{5mm}\noindent
This particle was called \cite{16} as $\cal L$-particle from the word
{\it lightest}. It should be emphasized that we predicted the
position of the new ``threshold" with a high accuracy. Of course, the
natural questions have been arisen.  What is the physical nature and
dynamical origin of $\cal L$-particle? Could $\cal L$-particle be
related to the experimentally observed diproton resonances spectrum?
We present the answers to these questions in the next section.

\section
{${\cal L}$-particle and Kaluza-Klein world}

The original idea of Kaluza and Klein is based on the hypothesis that
the input space-time is a $(4+d)$-dimensional space ${\cal
M}_{(4+d)}$ which can be represented as a tensor product of the
visible four-dimensional world $M_4$ with a compact internal
$d$-dimensional space $K_d$
\be
{\cal M}_{(4+d)} = M_4 \times K_d.   \label{1}
\ee
The compact internal space $K_d$ is space-like one i.e. it has only
spatial dimensions which may be considered as extra spatial
dimensions of $M_4$. In according with the tensor product structure 
of the space ${\cal M}_{(4+d)}$ the metric may be chosen in a
factorizable form. This means that if $z^M = 
\{ x^{\mu}, y^{m}\}$, ($M=0,1,\ldots,3+d,\, \mu = 0,1,2,3,\, m=1,2,
\ldots, d$), are local coordinates on ${\cal M}_{(4+d)}$ then the
factorizable metric looks like
\[
ds^{2} = {\cal G}_{MN}(z) dz^M dz^N =
g_{\mu \nu}(x) dx^{\mu} dx^{\nu} + \gamma_{mn}(x,y) dy^{m} dy^{n},
\]
where $g_{\mu \nu}(x)$ is the metric on $M_4$.

In the year 1921, Kaluza proposed a unification of the
Einstein gravity and the Maxwell theory of electromagnetism in four
dimensions starting from Einstein gravity in five dimensions. He
assumed that the five-dimensional space ${\cal M}_5$ had to be a
product of a four-dimensional space-time $M_4$ and a circle $S_1$:
${\cal M}_5 = M_4 \times S_1$. It was shown that the zero mode sector
of the Kaluza model is equivalent to
the four-dimensional theory which describes the Einstein gravity with
a four-dimensional general coordinate transformations and the Maxwell
theory of electromagnetism with a gauge transformations. 

Recently some models with extra dimensions have been proposed to
attack the electroweak quantum instability of the Standard Model
known as hierarchy problem between the electroweak and gravity
scales. It is obviously that the basic idea of the Kaluza-Klein
scenario may be applied to any model in Quantum Field Theory. As
example, let us consider the simplest case of (4+d)-dimensional model
of scalar field with the action  
\be
S = \int d^{4+d}z \sqrt{-{\cal G}} \left[
\frac{1}{2} \left( \partial_{M} \Phi \right)^2 - 
\frac{m^{2}}{2} \Phi^2 + \frac{G_{(4+d)}}{4!} \Phi^4
\right], 
\label{S}
\ee
where ${\cal G}=\det|{\cal G}_{MN}|$, ${\cal G}_{MN}$ is the metric
on ${\cal M}_{(4+d)} = M_4 \times K_d$, $M_4$ is pseudo-Euclidean
Minkowski space-time, $K_d$ is a compact internal $d$-dimensional
space with the characteristic size $R$. Let $\Delta_{K_{d}}$ be the
Laplace operator on the internal space $K_{d}$, and $Y_{n}(y)$ are
ortho-normalized eigenfunctions of the Laplace operator 
\be
\Delta_{K_{d}} Y_{n}(y) = -\frac{\lambda_{n}}{R^{2}} Y_{n}(y),  
\label{Yn}
\ee
and $n$ is a (multi)index labeling the eigenvalue
$\lambda_{n}$ of the eigenfunction $Y_{n}(y)$. A $d$-dimen\-sional
torus $T^{d}$ with equal radii $R$ is an especially simple example of
the compact internal space of extra dimensions $K_d$. The
eigenfunctions and eigenvalues in this special case look like 
\be
Y_n(y) = \frac{1}{\sqrt{V_d}} \exp \left(i \sum_{m=1}^{d}
n_{m}y^{m}/R
\right), \label{T}
\ee
\[
\lambda_n = |n|^2,\quad |n|^2= n_1^2 + n_2^2 + \ldots n_d^2, \quad
n=(n_1,n_2, \ldots, n_d),\quad -\infty \leq n_m \leq \infty,
\]
where $n_m$ are integer numbers, $V_d = (2\pi R)^d$ is the
volume of the torus.

To reduce the multidimensional theory to the effective
four-dimensional one we wright a harmonic expansion for
the multidimensional field $\Phi(z)$ 
\be
\Phi(z) = \Phi(x,y) = \sum_{n} \phi^{(n)}(x) Y_{n}(y). 
\label{H}
\ee
The coefficients $\phi^{(n)}(x)$ of the harmonic expansion
(\ref{H}) are called Kaluza-Klein (KK) excitations or KK modes, and
they usually include the zero-mode $\phi^{(0)}(x)$, corresponding to
$n=0$ and the eigenvalue $\lambda_{0} = 0$. Substitution of the KK
mode expansion into action (\ref{S}) and integration over the
internal space $K_{d}$ gives
\be
S = \int d^{4}x \sqrt{-g} \left\{  
\frac{1}{2} \left( \partial_{\mu} \phi^{(0)} \right)^{2} -
\frac{m^{2}}{2}
(\phi^{(0)})^{2} \right. + \frac{g}{4!} (\phi^{(0)})^{4} +
\ee
\[
+\left. \sum_{n \neq 0} \left[\frac{1}{2}
\left(\partial_{\mu} \phi^{(n)}
\right) 
\left(\partial^{\mu} \phi^{(n)} \right)^{*} -\frac 
{m_n^2}{2} \phi^{(n)}\phi^{(n)*} \right] 
+ \frac{g}{4!} (\phi^{(0)})^{2} \sum_{n\neq 0} \phi^{(n)}
\phi^{(n)*}\right\} + \ldots.  
\]
For the masses of the KK modes one obtains 
\be
m_{n}^{2} = m^{2} + \frac{\lambda_{n}}{R^{2}}, \label{m}
\ee
and the coupling constant $g$ of the four-dimensional theory is
related  to the coupling constant $G_{(4+d)}$ of the initial
multidimensional theory by the equation 
\be
  g = \frac{G_{(4+d)}}{V_d},  \label{g}
\ee
where $V_d$ is the volume of the compact internal space of extra
dimensions $K_d$. The fundamental coupling constant $G_{(4+d)}$ has
dimension $[\mbox{mass}]^{-d}$. So, the four-dimensional coupling
constant $g$ is dimensionless one as it should be.
Eqs.~(\ref{m},\ref{g}) represent the basic
relations of Kaluza-Klein scenario.  Similar relations take place for
other types of multidimensional quantum field theoretical models.
From four-dimensional point of view we can interpret each KK mode as
a particle with the mass $m_n$ given by Eq.~(\ref{m}). We see that in
according with Kaluza-Klein scenario any multidimensional
field contains an infinite set of KK modes, i.e. an infinite set of
four-dimensional particles with increasing masses, which is called
the Kaluza-Klein tower. Therefore, an experimental observation of
series KK excitations with a characteristic spectrum of the form
(\ref{m}) would be an evidence of the existence of extra dimensions.
So far the KK partners of the particles of the Standard Model have
not been observed. In the Kaluza-Klein scenario this fact can be
explained by a microscopic small size $R$ of extra dimensions
($R<10^{-17}\,cm$); in that case the KK excitations may be produced
only at super-high energies of the scale $E\sim 1/R > 1\,TeV$. Below
this scale only homogeneous zero modes with $n=0$ are accessible ones
for an observation in recent high energy experiments. That is why,
there is a hope to search the KK excitations at the future LHC and
other colliders. 

The most recent developments are related with a remarkable idea of
``brane world picture" according to which all matter fields (except
gravity) are localized on a three-dimensional submanifold -- brane --
embedded in fundamental multidimensional space. In the brane world
scenario extra dimensions may have large and even very large size.
Even though the models with the brane world scenario may rather seem
as exotic ones, nevertheless, they provide a base for a nontrivial
phenomenological issues related to the fundamental problems in
particle physics and cosmology. We refer with a pleasure the
interested reader to the excellent review articles \cite{17,18} and
many references therein.

In our recent paper \cite{19} we argued in favour of that the extra
dimensions have been observed for a long time in the experiments at
very low energies where the nucleon-nucleon dynamics has been
studied: We have shown that the structure of proton-proton total
cross section at very low energies revealed a clear signature of the
existence of the extra dimensions. Here we repeat our arguments
applying the main issues of Kaluza-Klein
approach to our concrete case. 

Let us assume that $\cal L$-particle
is related to the first KK excitation in the diproton system. Using
formula (\ref{m}) for the masses of KK modes, we can calculate the
scale (size) $R$ of the compact internal extra space. So, starting
from the formula
\be
\sqrt{s_{thr}}=2m_p+m_{\cal L}=2\sqrt{m_p^2+\frac{1}{R^2}},\label{RL}
\ee
one obtains
\be
\frac{1}{R}=\sqrt{m_{\cal L}(m_p+\frac{1}{4}m_{\cal L})}=41.481\,
MeV,\label{scale}
\ee
where $m_p=938.272\,MeV$ for the proton mass and Eq.~(\ref{L}) for
the
mass of $\cal L$-particle have been used. From Eq.~(\ref{scale}) it
follows
\be
R=24.1\,GeV^{-1}=4.75\,10^{-13}\mbox{cm}.\label{size}
\ee
It should be emphasized a remarkable fact: the size (\ref{size}) just
corresponds to the scale of distances where the strong Yukawa forces
in strength come down to the electromagnetic forces
$$g_{eff}=g_{\pi NN}\exp(-m_{\pi}R)\sim 0.5,\ \ \  (g^2_{\pi
NN}/4\pi=14.6).$$
On the other hand, for the fundamental mass scale  with account of
size (\ref{size}) in the case $d=6$  we find \cite{19}
\be
M\sim R^{-1}\left(\frac{M_{Pl}}{R^{-1}}\right)^{2/(d+2)}\mid_{d=6}\,
\sim 5\,\mbox{TeV}.\label{SM}
\ee
Mass scale (\ref{SM}) is just the scale accepted in the Standard
Model, and this is an interesting observation as well.

Going further on, let us build the Kaluza-Klein tower of KK
excitations by the formula\footnote{Similar formula has been
discussed in the literature \cite{28} but with a different physical
interpretation.} 
\be
M_n=2\sqrt{m_p^2+\frac{n^2}{R^2}},\quad (n=1,2,3,\ldots)\label{KK}
\ee
and compare it with the observed irregularities in the spectrum of
mass of the diproton system. The result of the comparison is shown in
Table 2. As it is seen from the Table 2, there is a quite remarkable
correspondence of the Kaluza-Klein picture with the
experiment. It is pleased for me to tell about it here at this
Conference because the main part from the list of diproton resonances
has been discovered and observed in Dubna.

Now, let us suppose that effective bosons $B_n$ with the masses
$m_n=n/R$ related to KK-excitations of a proton may have an effective
Yukawa-type interaction with the fermions
\be
L_{eff} = g_{eff}\bar\psi_f O\psi_f B_n,\label{Leff}
\ee
where $f$ denotes some fermion, for example lepton or quark. 

\begin{center}
Table 2. Kaluza-Klein tower of KK excitations of diproton system.

\vspace{5mm}
\begin{tabular}{|c|c|lc|}   \hline
n & $M_n(MeV)$ & $M_{exp}^{pp}(MeV)$ & Refs.\\ \hline
1 & 1878.38 & 1877.5 $\pm$ 0.5 & [20] \\ \hline
2 & 1883.87 & 1886 $\pm$ 1 & [14] \\ \hline
3 & 1892.98 & 1898 $\pm$ 1 & [14] \\ \hline
4 & 1905.66 & 1904 $\pm$ 2 & [21] \\ \hline 
5 & 1921.84 & 1916 $\pm$ 2 & [14] \\
  &         & 1926 $\pm$ 2 & [21] \\ \hline
  &         & 1937 $\pm$ 2 & [14] \\
  &         & 1942 $\pm$ 2 & [21] \\
6 & 1941.44 & 1945 $\pm$ 2.5 & [15] \\
  &         & 1955 $\pm$ 2 & [15] \\ 
  &         & 1956 $\pm$ 3 & [22] \\\hline
7 & 1964.35 & 1965 $\pm$ 2 & [14] \\
  &         & 1969 $\pm$ 2 & [24] \\ \hline 
8 & 1990.46 & 1980 $\pm$ 2 & [14] \\
  &         & 1999 $\pm$ 2 & [14] \\ \hline
9 & 2019.63 & 2017 $\pm$ 3 & [14] \\ \hline
  &         & 2035 $\pm$ 8 & [22] \\
10 & 2051.75 & 2046 $\pm$ 3 & [14] \\
   &         & 2050 $\pm$ 3.2 & [25] \\ \hline
11 & 2086.68 & 2087 $\pm$ 3 & [14] \\ \hline
   &         & 2120 $\pm$ 3.2 & [25] \\
12 & 2124.27 & 2121 $\pm$ 3 & [26] \\
   &         & 2129 $\pm$ 5 & [14] \\ \hline
   &         & 2140 $\pm$ 9 & [22] \\
13 & 2164.39 & 2150 $\pm$ 12.6 & [25] \\
   &         & 2172 $\pm$ 5 & [14] \\ \hline
   &         & 2192 $\pm$ 3 & [26] \\
14 & 2206.91 & 2217 & [16] \\
   &         & 2220 & [23] \\ \hline
15 & 2251.67 & 2238 $\pm$ 3 & [14] \\   
   &         & 2240 $\pm$ 5 & [26] \\ \hline
16 & 2298.57 & 2282 $\pm$ 4 & [14,27] \\ \hline
17 & 2347.45 & 2350 & [23] \\ \hline
\end{tabular}
\end{center}
\vspace{5mm}
If $m_n>2m_f$ then effective bosons may decay into
fermion-antifrermion pair. For the partial width of such decay in the
lowest order over coupling constant we have
\be
\Gamma_n = \frac{\alpha_{eff}m_n}{2}F_O(x^2_n),\label{Gamma}
\ee
where $\alpha_{eff}=g^2_{eff}/4\pi$, $x^2_n=m^2_f/m^2_n$,
$F_O(x^2)=(1-4x^2)^{3/2}$ for $O=1$ and $F_O(x^2)=(1-4x^2)^{1/2}$ for
$O=\gamma_5$. In that case one obtains an estimation
\be
\Gamma_n \sim n\cdot 0.4\,\mbox{MeV}.\label{Gamma2}
\ee

It's clear from the physics under consideration that a life time of
the diproton resonances will be defined by the decays of effective
bosons $B_n$. This is a remarkable fact that crude estimation
(\ref{Gamma2}) is in a good agreement with an experiment and gives an
explanation of (super)narrowness of dibaryons peaks. Moreover,
estimation (\ref{Gamma2}) shows that the larger the dibaryon mass is,
the larger is the width of the dibaryon.   

Here we have concerned the simplest model where the protons were
considered as a scalar particles. It is well known that account of
fermionic degrees of freedom may result the nontrivial problems
related to both the index and the kernel of Dirac operator on a
generic compact manifold. However, since the kernel of Dirac operator
is equal to the kernel of its square, we can say with confidence that
account of fermionic degrees of freedom for a proton will not change
our main conclusion. 

Certainly, we have also considered here the simplest case of
Kaluza-Klein picture: The built KK-tower corresponds to either
one-dimensional compact extra space or d-dimensional equal radii
torus with the constraint
\be
n = \sqrt{n_1^2 + n_2^2 + \ldots n_d^2}=1,2,3,\ldots,\label{Dio}
\ee
where $n_i(i=1,\ldots,d)$ are integer numbers. The constraint
(\ref{Dio}) corresponds to the special (Diophantus!) selection of the
states. It's clear that in general case of generic extra compact
manifold we would have a significantly more wealthy spectrum of
KK-excitations. One could imagine that there exist such extra compact
manifold with a suitable geometry where KK-excitations of a few input
fundamental entities (proton, electron, photon, etc.) would provide
the experimentally observed spectrum of all particles, their
resonances and nuclei states. As we hope, it would be possible to
find in this way the global solution of the Spectral Problem. Anyhow,
we believe that such perfect extra compact manifold with a beautiful
geometry and its good-looking shapes exist.

\section{Conclusion}

Investigating the three-body forces open many new pages in the study
of fundamental dynamics of particles and nuclei:
\begin{itemize}
\item
{\bf Three-body forces define the dynamics of one-par\-ticle
inclusive
reactions}.
\item
{\bf We have to take into account a contribution of three-body
forces in scattering from deuteron and, in general, from nuclei}.
\item
{\bf New scaling characteristics in shadow dynamics in scattering
from deute\-ron have been established by account of three-body
forces}.
\item
{\bf Introduction of three-body forces resulted the discovery of
global structure of (anti)proton-proton total cross sections}.
\item
{\bf Investigating the three-body forces allowed us to predict a new
particle ($\cal L$-particle), describing a new scale of internucleon
distances, where strong Yukawa forces compared with
electromagnetic ones}.
\end{itemize}
\noindent
We would like to emphasize that multidimensional (six-dimensional)
space is a natural space to describe the properties of three-body
forces. Geniusly simple formula provided by Kaluza-Klein approach so
accurately described the mass spectrum of diproton system, and
certainly it was not an accidental coincidence. This means that {\bf
the existence of the extra dimensions was experimentally proved in
the experiments  at very low energies where the nucleon-nucleon
dynamics had been studied, but we did not understand it. However, now
it seems we understand it}. 

\section*{Acknowledgements}
It is my great pleasure to express thanks to the Organizing 
Committee for the kind invitation to attend the XXXII International
Symposium on Multiparticle Dynamics. I would like to especially thank
A.N. Sissakian, G. Kozlov and all local organizers from Dubna for
excellently organized Conference.

\newpage 
\vspace*{20mm}
\begin{figure}[ht]
\begin{center}
\begin{picture}(298,184)
\put(20,10){\includegraphics{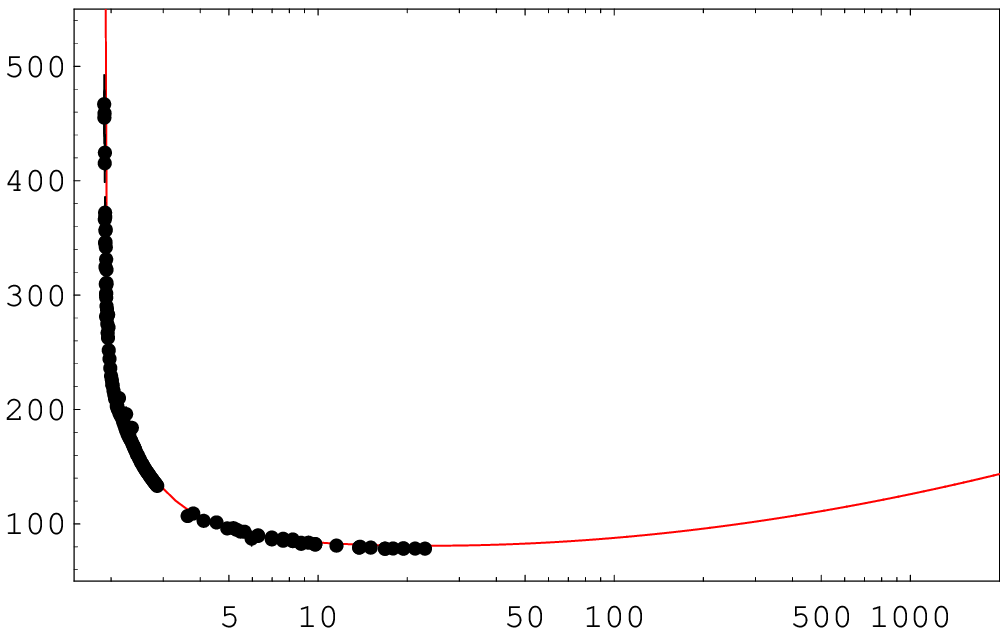}}
\put(155,0){$\sqrt{s}\, (GeV)$}
\put(0,90){\rotatebox{90}{$\sigma^{tot}_{\bar{p}d} (mb)$}}
\end{picture}
\end{center}
\caption{The total antiproton-deuteron cross-section compared with
the theory. Statistical and systematic errors added in
quadrature.}\label{fig1}
\end{figure}

\vspace*{10mm}
\begin{figure}[ht]
\begin{center}
\begin{picture}(288,184)
\put(15,10){\includegraphics{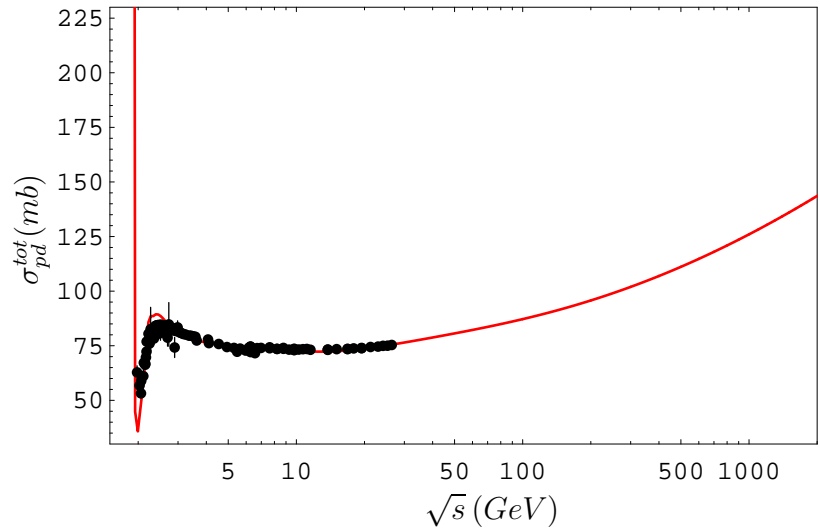}}
\put(155,0){$\sqrt{s}\, (GeV)$}
\put(0,88){\rotatebox{90}{$\sigma^{tot}_{pd} (mb)$}}
\end{picture}
\end{center}
\caption{The total proton-deuteron cross-section compared with the
theory without any free parameters. Statistical and systematic errors
added in quadrature.}\label{fig2}
\end{figure}

\newpage
\begin{figure}[t]
\begin{center}
\begin{picture}(288,194)
\put(15,10){\includegraphics*[]{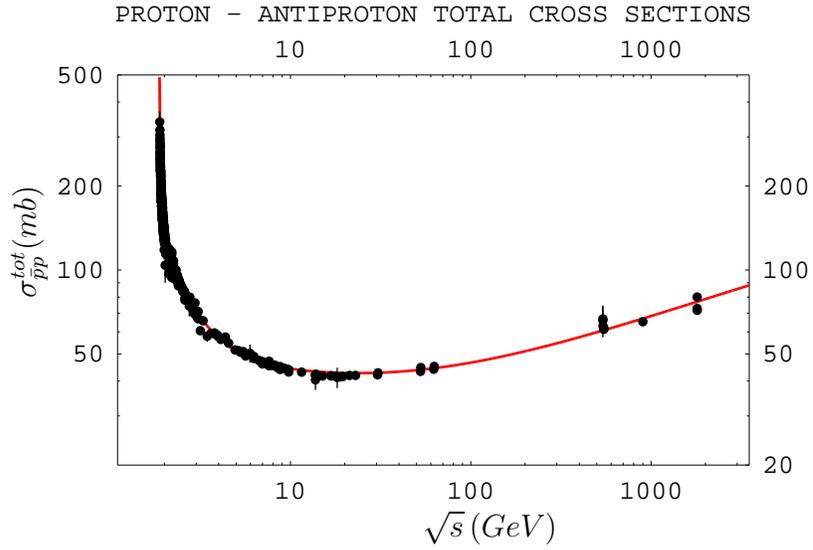}}
\put(155,0){$\sqrt{s}\, (GeV)$}
\put(0,90){\rotatebox{90}{$\sigma^{tot}_{\bar{p}p} (mb)$}}
\end{picture}
\end{center}
\vspace*{5mm}
\caption[]{\protect
{The proton-antiproton total cross sections
versus $\sqrt{s}$ compared with the theory. Solid line represents our
fit to the data. Statistical and systematic errors added in
quadrature.}}
\label{fig:3}
\end{figure}

\begin{figure}[ht]
\begin{center}
\begin{picture}(288,200)
\put(15,10){\includegraphics{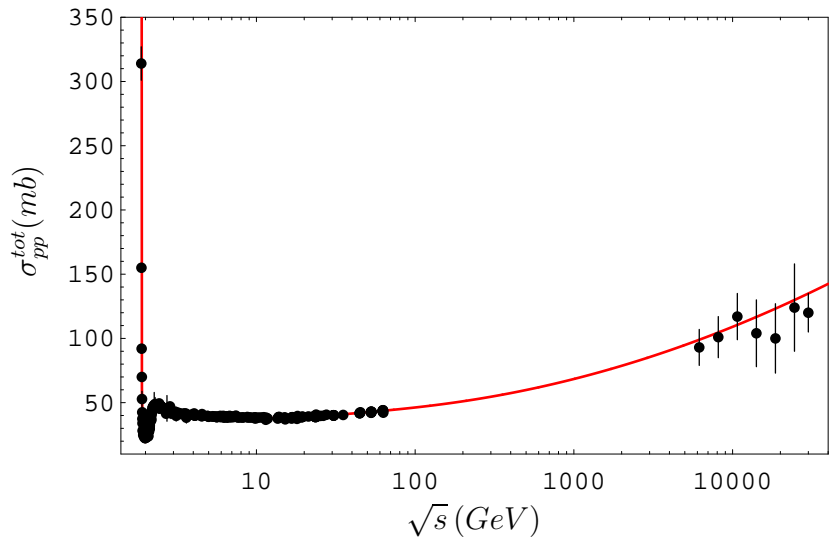}}
\put(144,0){$\sqrt{s}\, (GeV)$}
\put(-5,95){\rotatebox{90}{$\sigma^{tot}_{pp} (mb)$}}
\end{picture}
\end{center}
\caption[]{\protect{The proton-proton total cross-section versus
$\sqrt{s}$ with the cosmic rays data points from Akeno Observatory
and Fly's Eye Collaboration. Solid line corresponds to our theory
predictions.}}
\label{fig:4}
\end{figure}
\newpage

\begin{figure}[t]
\begin{center}
\begin{picture}(288,188)
\put(15,10){\includegraphics{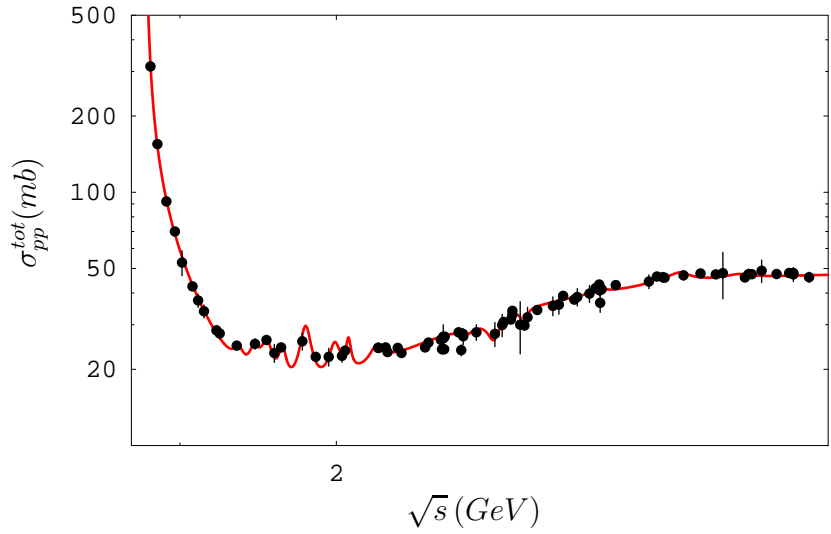}}
\put(144,0){$\sqrt{s}\, (GeV)$}
\put(-5,95){\rotatebox{90}{$\sigma^{tot}_{pp} (mb)$}}
\end{picture}
\end{center}
\caption{\protect{The proton-proton total cross-section versus
$\sqrt{s}$ at low energies. Solid line corresponds to our theory
predictions.}}
\label{fig:5}
\end{figure}

\begin{figure}[ht]
\begin{center}
\begin{picture}(288,188)
\put(15,10){\includegraphics{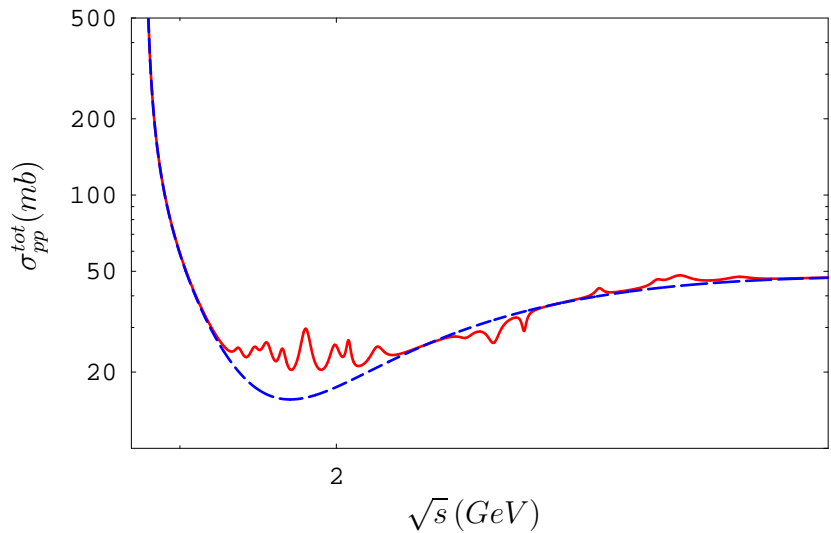}}
\put(144,0){$\sqrt{s}\, (GeV)$}
\put(-5,95){\rotatebox{90}{$\sigma^{tot}_{pp} (mb)$}}
\end{picture}
\end{center}
\caption{\protect{The resonance structure for the proton-proton
total cross-section versus $\sqrt{s}$ at low energies. Solid line is
our theory predictions. Dashed line corresponds to the ``background"
where all resonances are switched off.}}
\label{fig:Fig.6}
\end{figure}

\end{document}